\documentstyle[epsfig]{elsart}
\def\bm#1{\mbox{\boldmath$#1$}} 
\begin{document}
\begin{frontmatter}
\title{$0^+ \to 2^+$ $0\nu\beta\beta$ decay triggered directly by
the Majorana neutrino mass}
\author{T. Tomoda\thanksref{email}}
\address{Aomori University, 2-3-1 Kobata, Aomori 030-0943, Japan}
\address{and}
\address{The Institute of Physical and Chemical Research (RIKEN),
2-1 Hirosawa, Wako 351-0198, Japan}
\thanks[email]{E-mail: tomoda@aomori-u.ac.jp}
\begin{abstract}
We treat $0^+ \to 2^+$ $0\nu\beta\beta$ decays taking into account
recoil corrections to the nuclear currents. The decay probability can be
written as a quadratic form of the effective coupling constants of the
right-handed leptonic currents {\it and\/} the effective neutrino mass.
We calculate the nuclear matrix elements for the
$0^+ \to 2^+_1$ $0\nu\beta\beta$ decays of $^{76}$Ge and $^{100}$Mo, and
demonstrate that the {\it relative\/} sensitivities of $0^+ \to 2^+$
decays to the neutrino mass and the right-handed currents are comparable
to those of $0^+ \to 0^+$ decays.
\end{abstract}
\end{frontmatter}
The neutrinoless double beta ($0\nu\beta\beta$) decay can take place
through an exchange of neutrino between two quarks in nulei if the
electron neutrino is a Majorana particle and has a nonvanishing mass
and/or right-handed couplings \cite{HS,DKT,TomR}. There may be
other possible mechanisms such as those involving supersymmetric
particles which also cause the decay of two neutrons into two protons
and two electrons \cite{Moh,Ver,HKK}. In the present work, however,
we restrict ourselves to the conventional two-nucleon and $\Delta$
mechanisms of $0\nu\beta\beta$ decay through light Majorana neutrino
exchange. From the analises of experimental data on $0^+ \to 0^+$
$0\nu\beta\beta$ decays, stringent limits on the effective neutrino
mass and the effective coupling constants of the right-handed leptonic
currents have been deduced (see {\it e.g.} \cite{TomR,Mor} and the
references quoted therein). On the other hand it still seems to be
believed widely that $0^+ \to 2^+$ $0\nu\beta\beta$ decays are sensitive
only to the right-handed currents. In view of the theorem that the
electron neutrino should have a nonvanishing Majorana mass if
$0\nu\beta\beta$ decay occurs anyway \cite{SV,Nie,Tak}, an observation
of $0\nu\beta\beta$ decay due to right-handed interactions would
certainly mean also a nonvanishing Majorana mass of the electron
neutrino. The purpose of the present work is, however, not to
investigate the role of the Majorana neutrino mass in such a sense,
but to demonstrate that it causes $0^+ \to 2^+$ $0\nu\beta\beta$ decays
directly.

A direct contribution of the neutrino mass to $0^+ \to 2^+$
$0\nu\beta\beta$ decays was considered in \cite{TomR} taking into
account the nuclear recoil currents, and the inverse half-life was
given as
\begin{eqnarray}
[\tau_{1/2}^{0\nu}(0^+\to 2^+)]^{-1}&=&
F_{1+}(Z_{1+})^2+F_{1-}(Z_{1-})^2+F_{2+}(Z_{2+})^2+F_{2-}(Z_{2-})^2\, ,
\nonumber\\
&&
\end{eqnarray}
where $F_{j\pm}$ ($j=1, 2$) are the phase space integrals and
\begin{eqnarray}
Z_{1\pm}&=&M_\lambda \langle\lambda\rangle
 -M_\eta \langle\eta\rangle
 \pm M_{\rm m} \frac{\langle m_{\nu}\rangle}{m_{\rm e}}\, ,\nonumber\\
Z_{2\pm}&=&M'_\eta \langle\eta\rangle
 \pm M_{\rm m} \frac{\langle m_{\nu}\rangle}{m_{\rm e}}\, ,
\label{eq:z}
\end{eqnarray}
with the electron mass $m_{\rm e}$ and
\begin{eqnarray}
\langle m_{\nu}\rangle &=& {\sum_j}' U_{ej}^2m_j\, ,\nonumber\\
\langle\lambda\rangle &=& \lambda{\sum_j}' U_{ej}V_{ej}\, ,\nonumber\\
\langle\eta\rangle &=& \eta{\sum_j}' U_{ej}V_{ej}\, .
\end{eqnarray}
Here $m_j$ is the mass of the eigenstate Majorana neutrino $N_j$.
$U_{ej}$ and $V_{ej}$ are the amplitudes of $N_j$ in the left- and
right-handed electron neutrinos, $\lambda$ and $\eta$ the coupling
constants of the right-handed leptonic current with the right- and
left-handed hadronic currents, and the summations should be taken over
light neutrinos ($m_j \ll 100$\ MeV). The nuclear matrix elements
$M_\alpha$ ($\alpha=\lambda$, $\eta$, m) are defined by
\begin{equation}
M_\alpha = \langle 2^+_{\rm F}\|{\textstyle\frac{1}{2}}\sum_{n,m}
\tau^+_n\tau^+_m(\bm{M}_\alpha)_{nm}\|0^+_{\rm I}\rangle\, .
\label{eq:mal}
\end{equation}
The explicit forms of the two body operators $\bm{M}_\lambda$,
$\bm{M}_\eta$ and $\bm{M}_\eta'$ were given in \cite{TomN} including
the contribution of the $\Delta$ mechanism, in which the
$0\nu\beta\beta$ decay proceed through an exchange of a Majorana
neutrino between two quarks in the same baryon in a nucleus. On the
other hand the operator $\bm{M}_{\rm m}$ was derived in \cite{TomR} as
\begin{eqnarray}
(\bm{M}_{\rm m})_{nm} & = & -{\textstyle\frac{1}{2}}im_{\rm e}\Big\{
[\bm{r}_{nm}\otimes(\bm{\sigma}_nC_m-\bm{\sigma}_mC_n)]^{(2)}\nonumber\\
& & +i(g_{\rm V}/g_{\rm A})[\bm{r}_{nm}\otimes
(\bm{D}_n\times\bm{\sigma}_m-\bm{D}_m\times\bm{\sigma}_n)]^{(2)}
\nonumber\\
& & +(g_{\rm V}/g_{\rm A})^2
[\bm{r}_{nm}\otimes(\bm{D}_n-\bm{D}_m)]^{(2)}\Big\}H(r_{nm})\, ,
\label{eq:mm}
\end{eqnarray}
where $\bm{r}_{nm} = \bm{r}_{n} - \bm{r}_{m}$, $H(r)$ is the neutrino
propagation function, $g_{\rm V}$ and $g_{\rm A}$ the vector and axial
vector coupling constants. $C_n$ and $\bm{D}_n$ are the recoil
correction terms to the axial vector and vector nuclear currents
\cite{DKT,TFSG} given by
\begin{eqnarray}
C_n &=& (\bm{p}_n+\bm{p}'_n)\cdot \bm{\sigma}_n/2M\, ,\nonumber\\
\bm{D}_n &=& 
[\bm{p}_n+\bm{p}'_n-i\mu_\beta\bm{\sigma}_n\times(\bm{p}_n-\bm{p}'_n)]
/2M\, ,
\end{eqnarray}
where $\bm{p}_n$ and $\bm{p}'_n$ are the initial and final nucleon
momenta, $M$ the nucleon mass, and $\mu_\beta = 4.7$. The above
expression for $\bm{M}_{\rm m}$ is, however, not suitable for numerical
calculations as it stands. Therefore, as was done for $\bm{M}_\lambda$,
$\bm{M}_\eta$ and $\bm{M}_\eta'$ in \cite{TomN}, we expand it in terms
of the operators $\bm{M}_{inm}$ with simpler spin and orbital
structures,
\begin{equation}
(\bm{M}_{\rm m})_{nm} = {\sum_i}C_{{\rm m}i}\bm{M}_{inm}\, .
\end{equation}
We define the matrix element $M_i$ of the operator $\bm{M}_{inm}$
analogously to Eq.\ (\ref{eq:mal}). The coefficients $C_{{\rm m}i}$
and the two-body operators $\bm{M}_{inm}$ are listed in Table 1, where
\begin{equation}
\begin{array}[b]{ll}
h = r_{nm}H(r_{nm})\, , &
 h' = -r_{nm}H'(r_{nm})\, ,\\
\bm{S}_{\lambda nm} =
 [\bm{\sigma}_n\otimes\bm{\sigma}_m]^{(\lambda)}\, , &
 \bm{S}_{\pm nm} = \bm{\sigma}_n \pm \bm{\sigma}_m\, ,\\
\bm{y}_{Knm} = [\hat{\bm{r}}_{nm}\otimes\hat{\bm{r}}_{nm}]^{(K)}\, , &
 \bm{Y}_{Knm} = [\hat{\bm{r}}_{nm}\otimes\hat{\bm{r}}_{+nm}]^{(K)}\,
 (r_{+nm}/r_{nm})\, ,\\
\bm{y}'_{Knm} = i[\hat{\bm{r}}_{nm}\otimes\bm{p}_{nm}]^{(K)}\, ,
 \hspace*{3ex}&
 \bm{Y}'_{Knm} = i[\hat{\bm{r}}_{nm}\otimes\bm{P}_{nm}]^{(K)}\, ,\\
\bm{r}_{+nm}  = \bm{r}_{n} + \bm{r}_{m}\, , &
 \hat{\bm{a}} = \bm{a}/|\bm{a}|\, ,\\
\bm{p}_{nm}   = {\textstyle\frac{1}{2}}(\bm{p}_{n} - \bm{p}_{m})\, , &
 \bm{P}_{nm}   = \bm{p}_{n} + \bm{p}_{m}\, .\\
\end{array}
\end{equation}

As was described in detail in \cite{TomN}, $\bm{M}_\lambda$ and
$\bm{M}_\eta$ can be expanded in terms of $\bm{M}_{inm}$ with
$1 \le i \le 5,\, 8 \le i \le 13$, and $\bm{M}_\eta'$ in terms of
$\bm{M}_{inm}$with $i = 6,\, 7$ (for the definition of $\bm{M}_{inm}$
with $6 \le i \le 13$, which do not appear in Table 1, see
\cite{TomN}). Of these operators, $\bm{M}_{inm}$ with $8 \le i \le 13$
are related to the $0\nu\beta\beta$ transitions which involve virtual
$\Delta$ particles in nuclei, and they are induced by the operator
$\bm{M}_{2nm}$ interpreted as acting on two quarks in a nucleon or a
$\Delta$ particle.

The new operators $\bm{M}_{inm}$ with $14 \le i \le 25$ appear only in
the expansion of $\bm{M}_{\rm m}$. In the derivation of $C_{{\rm m}i}$
listed in Table 1, we have not taken into account the $\Delta$
mechanism yet. Under the same assumption of the non-relativistic
constituent quark model about the $\Delta$ mechanism as was made in
\cite{TomN}, the operators in Table 1 except $\bm{M}_{inm}$ with
$i = 2,\, 16,\, 17$ do not contribute when interpreted as acting on two
quarks in a nucleon or a $\Delta$ particle. Since the relation
$\mu_\beta = g_{\rm V} = g_{\rm A} = 1$ holds for the quark currents,
we see $C_{{\rm m}i} = 0$ for $i = 2,\, 16$. The only possible
contribution of $\bm{M}_{17nm}$ to $\bm{M}_{\rm m}$ is estimated to be
about $m_{\rm e}/2M$ of the $\Delta$ mechanism contributions to
$\bm{M}_\lambda$ and $\bm{M}_\eta$. Therefore we will neglect the
$\Delta$ mechanism for the calculation of $\bm{M}_{\rm m}$ in the
present work.

We calculate the nuclear matrix elements for $0^+_1 \to 2^+_1$
$0\nu\beta\beta$ decay of $^{76}$Ge and $^{100}$Mo using the method
given in \cite{TomN}. We describe the initial $0^+_1$ and final $2^+_1$
nuclear states in terms of the Hartree-Fock-Bogoliubov type wave
functions which are obtained by variation after particle-number and
angular-momentum projection \cite{TomN,TFSG,SGF}. For the case of
$^{76}$Ge decay, the calculation of the matrix elements $M_i$ with
$1 \le i \le 13$ has been performed in \cite{TomN}. In the present work
we calculate only the new ones with $14 \le i \le 25$ using the nuclear
wave functions obtained in \cite{TomN}. In order to calculate all $M_i$
with $1 \le i \le 25$ for the $^{100}$Mo decay, the nuclear wave
functions of $^{100}$Mo$(0_1^+)$ and $^{100}$Ru$(2_1^+)$ are constructed
in the same manner as in the case of the $^{76}$Ge decay. Table 2 shows
the calculated matrix elements $M_{\rm m}$ for the $^{76}$Ge and
$^{100}$Mo decays as a sum of the products $C_{{\rm m}i}M_i$. It should
be noted that the matrix elements of the operators with rank 0 spin
part, {\it i.e.} $M_1$, $M_4$, $M_{14}$ and $M_{15}$ have the dominant
contributions to $M_{\rm m}$. Table 3 summarizes the calculated matrix
elements $M_\lambda$, $M_\eta$, $M'_\eta$ and $M_{\rm m}$ for the
$^{76}$Ge and $^{100}$Mo decays. 

The differential rate for $0^+ \to 2^+$ $0\nu\beta\beta$ decay with the
energy of one of the emitted electrons $\epsilon_1$ and the angle
between the two electrons $\theta_{12}$ can be written as
\begin{equation}
\frac{{\rm d}^2W_{0\nu}}{{\rm d}\epsilon_1{\rm d}\cos\theta_{12}} =
 a^{(0)}(\epsilon_1) + a^{(1)}(\epsilon_1){\rm P}_1(\cos\theta_{12})
 + a^{(2)}(\epsilon_1){\rm P}_2(\cos\theta_{12})\, .
\end{equation}
Each of the angular correlation coefficients $a^{(k)}(\epsilon_1)$
$(k = 0, 1, 2)$ can be expressed as a sum of the products of an
electron phase space factor and a second order monomial of $Z_{j\pm}$
defined in Eq.\ (\ref{eq:z}). The explicit form of
$a^{(0)}(\epsilon_1)$, which yields $(\ln 2)/2$ times the right hand
side of Eq.\ (1) upon integration over $\epsilon_1$, can be readily
obtained by combining the relevant equations in \cite{TomR}. Since the
expressions for $a^{(1)}(\epsilon_1)$ and $a^{(2)}(\epsilon_1)$ are
rather complicated, they will be given elsewhere. Numerical calculations
show that $a^{(1)}(\epsilon_1)$ is dominated by a term with the factor
$-(Z_{1+})^2 + Z_{2+}Z_{2-}$ times a positive function of $\epsilon_1$,
whereas $a^{(2)}(\epsilon_1)$ by a term with the factor
$2Z_{1+}Z_{1-} - (Z_{2+})^2 - (Z_{2-})^2$. For later reference we denote
these two factors as $z^{(1)}$ and $z^{(2)}$, respectively.

Figure 1 shows the single electron spectra
${\rm d}W_{0\nu}/{\rm d}\epsilon_1 = 2a^{(0)}$ and the ratios of the
angular correlation coefficients $a^{(1)}/a^{(0)}$ and $a^{(2)}/a^{(0)}$
for the three limiting cases, (a) $\langle\lambda\rangle \ne 0$,
(b) $\langle\eta\rangle \ne 0$ and (c) $\langle m_\nu\rangle \ne 0$.
Since the coefficients $a^{(k)}(\epsilon_1)$ depend on the parameters
$\langle\lambda\rangle$, $\langle\eta\rangle$ and $\langle m_\nu\rangle$
through $Z_{j\pm}$, the results shown in Fig.\ 1 are independent of
nuclear models for the cases (a) and (c). We can also easily understand
the signs of $a^{(1)}$ and $a^{(2)}$ from the relations
$z^{(1)} = -(M_\lambda\langle\lambda\rangle)^2$ and
$z^{(2)} = 2(M_\lambda\langle\lambda\rangle)^2$ for the case (a), and
$z^{(1)} = -2(M_{\rm m}\langle m_\nu\rangle/m_{\rm e})^2$ and
$z^{(2)} = -4(M_{\rm m}\langle m_\nu\rangle/m_{\rm e})^2$ for the case
(c). On the other hand for the case (b), we obtain
$z^{(1)}= -(M_\eta\langle\eta\rangle)^2 + (M'_\eta\langle\eta\rangle)^2$
 and
$z^{(2)}=2(M_\eta\langle\eta\rangle)^2 -2(M'_\eta\langle\eta\rangle)^2$,
and consequently a cancellation between the contributions of $M_\eta$
and $M'_\eta$ occurs when these are of comparable magnitudes. This is
just the case for the $^{100}$Mo decay, but not for the $^{76}$Ge decay
where $M'_\eta$ is much smaller than $M_\eta$ so that there is no
significant difference between the cases (a) and (b) in the angular
correlation. It should also be noted in Fig.\ 1 that the single electron
spectra for all the three cases (a), (b) and (c) have approximately the
same shape. This is in contrast with the $0^+ \to 0^+$ decays where the
spectrum for $\langle\lambda\rangle \ne 0$ is very different from those
for $\langle m_\nu\rangle \ne 0$ or $\langle\eta\rangle \ne 0$
\cite{DKT,TomR}.

Using the matrix elements in Table 3 and the phase space integrals
$F_{j\pm}$ calculated in \cite{TomR}, we can deduce from the
experimental data
$\tau_{1/2}^{0\nu}(0^+\to 2_1^+)$ $>8.2 \times 10^{23}$ yr (90\% C.L.)
\cite{Mai} for the $^{76}$Ge decay the constraints on the right-handed
current couplings and the effective neutrino mass listed in Table 4. As
for the $^{100}$Mo decay, the Osaka group has obtained the limit
$\tau_{1/2}^{0\nu}(0^+\to 2_1^+)$ $>1.4 \times 10^{22}$ yr (68\% C.L.)
\cite{Kud} assuming $\langle\lambda\rangle \ne 0$. Because of the
differences in the angular correlation as we see from Fig.\ 1, an
analysis of the same raw experimental data might yield a half-life limit
significantly different from the above value especially for the case
$\langle\eta\rangle \ne 0$. However we assume here just the same
half-life limit also for the cases $\langle\eta\rangle \ne 0$ and
$\langle m_\nu\rangle \ne 0$ in order to compare the resulting
constraints with those from the $^{76}$Ge data.

The limits which can be deduced from the experimental bound
$\tau_{1/2}^{0\nu}(0^+\to 0^+) >5.7 \times 10^{25}$ yr (90\% C.L.)
\cite{Bau} on the $0^+ \to 0^+$ decay of $^{76}$Ge using the nuclear
matrix elements of \cite{TF} are
$|\langle\lambda\rangle|<3.8\times 10^{-7}$,
$|\langle\eta\rangle|<2.2\times 10^{-9}$ and
$|\langle m_\nu\rangle|<0.19$ eV.
Comparing these limits with those of Table 4, we notice the considerable
difference in the {\it absolute\/} sensitivities between the
$0^+ \to 0^+$ and $0^+ \to 2^+$ decays, which reflects the smaller
$Q$-value as well as the higher electron partial waves associated with
the latter. However, it should be stressed here that the
{\it relative\/} sensitivities to $\langle m_\nu\rangle$ and
$\langle\eta\rangle$ are comparable in both cases. In other words,
$\langle m_\nu\rangle=1$ eV would give roughly the same decay rate as
$\langle\eta\rangle = 10^{-8}$ in the $0^+ \to 2^+$ as well as in the
$0^+ \to 0^+$ decays. At the same time it should also be noted that the
$0^+ \to 2^+$ decay is relatively more sensitive to
$\langle\lambda\rangle$.

In summary, we have calculated $0^+ \to 2^+$ $0\nu\beta\beta$ decay
rates taking into account the recoil corrections to the nuclear
currents. As a result, the expression for the decay probability becomes
a quadratic form of not only the effective coupling constants
$\langle\lambda\rangle$ and $\langle\eta\rangle$ of the right-handed
leptonic currents but also the effective neutrino mass
$\langle m_\nu\rangle$ which would be totally absent without the
inclusion of the recoil corrections. In other words, the recoil
corrections give the {\it lowest\/} order contribution to the
$0^+ \to 2^+$ $0\nu\beta\beta$ decay for the case where
$\langle\lambda\rangle = \langle\eta\rangle = 0$ and
$\langle m_\nu\rangle \ne 0$. Furthermore, by the numerical calculation
of the relevant nuclear matrix elements, we have demonstrated that the
{\it relative\/} sensitivities of $0^+ \to 2^+$ decays to
$\langle m_\nu\rangle$ and $\langle\eta\rangle$ are comparable to those
of $0^+ \to 0^+$ decays.

The author thanks S.\ Yamaji at RIKEN for his warm hospitality.

\begin{figure}
\centering
\epsfig{file=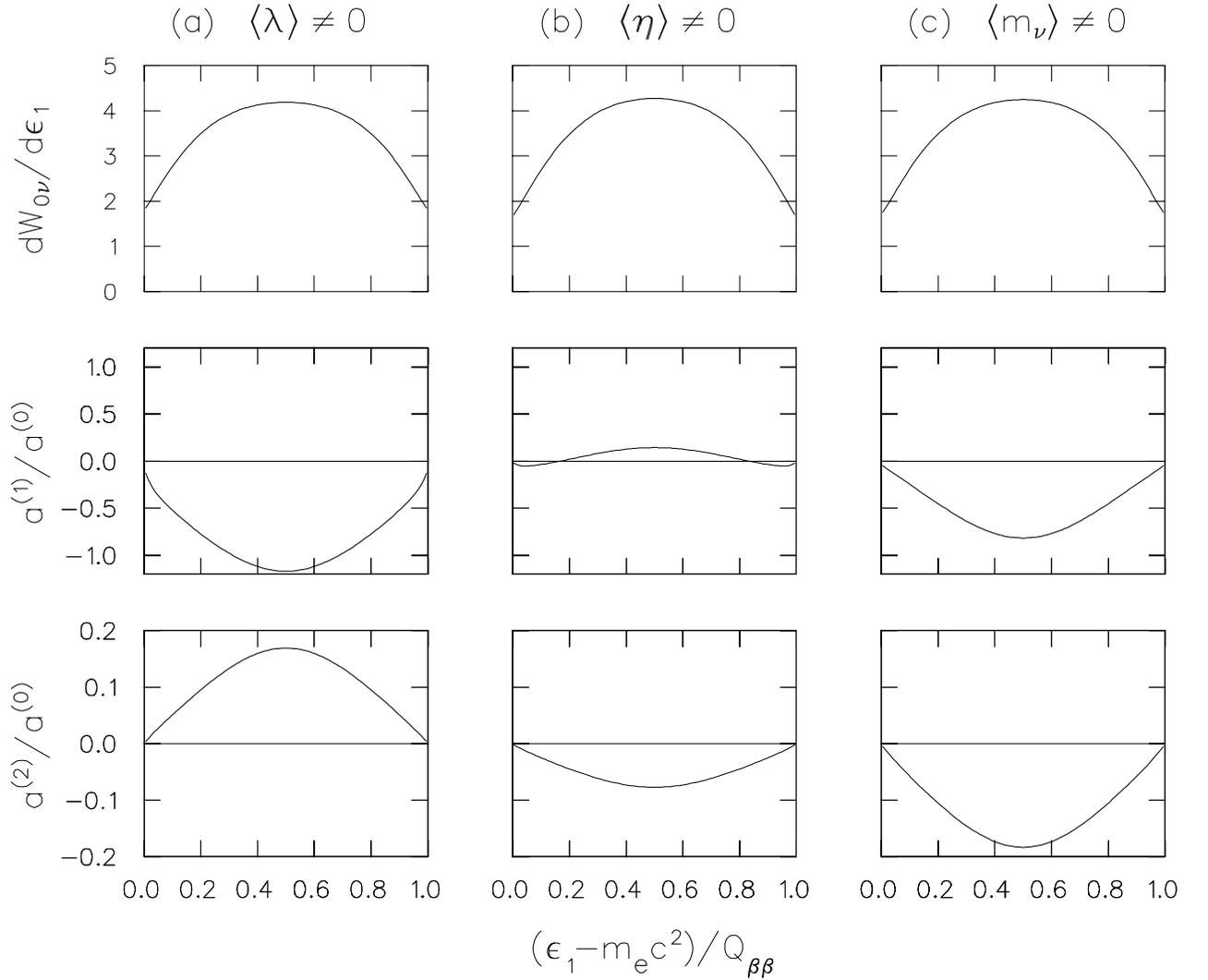,width=14cm,angle=90}
\caption{Single electron spectrum
${\rm d}W_{0\nu}/{\rm d}\epsilon_1$ in arbitrary units and the ratios of
the angular correlation coefficients $a^{(1)}/a^{(0)}$ and
$a^{(2)}/a^{(0)}$ for the $0^+ \to 2^+_1$ $0\nu\beta\beta$ decay of
$^{100}$Mo. They are all plotted against the kinetic energy fraction of
one of the two emitted electrons, where
$Q_{\beta\beta}(0^+ \to 2^+_1) = 2.494$ MeV. Only one of the three
lepton number violating parameters is assumed to be nonvanishing for
each of the three cases: (a) $\langle\lambda\rangle \ne 0$,
(b) $\langle\eta\rangle \ne 0$ and (c) $\langle m_\nu\rangle \ne 0$.}
\end{figure}
\begin{table}[h] 
\caption{The operators $\bm{M}_{inm}$ and the coefficients
$C_{{\rm m}i}\,$, the latter in units of the electron-nucleon mass ratio
$m_e/M$.}
\medskip
\begin{tabular}{rlc}
\hline
$i$ & $\bm{M}_{inm}$ & $C_{{\rm m}i}$\\
\hline
1 & $-\sqrt{3}h'\bm{S}_0\bm{y}_2$
  & $-\frac{1}{3}[\mu_{\beta}(g_{\rm V}/g_{\rm A}) + \frac{1}{2}]$\\
2 & $h'\bm{S}_2$
  & $\frac{1}{6}[\mu_{\beta}(g_{\rm V}/g_{\rm A}) - 1]$\\
3 & $h'[\bm{S}_2\otimes\bm{y}_2]^{(2)}$
  & $-\frac{\sqrt{7}}{4\sqrt{3}}[\mu_{\beta}(g_{\rm V}/g_{\rm A}) -1]$\\
4 & $h'\bm{y}_2$ & $\frac{1}{2}(g_{\rm V}/g_{\rm A})^2$\\
5 & $h'[\bm{S}_+\otimes\bm{y}_2]^{(2)}$
  & $\frac{\sqrt{3}}{4\sqrt{2}}
     [\mu_{\beta}(g_{\rm V}/g_{\rm A})^2  - (g_{\rm V}/g_{\rm A})]$\\
14 & $-\sqrt{3}h\bm{S}_0\bm{y}'_2$ & $\frac{1}{3}$\\
15 & $h\bm{y}'_2$ & $-(g_{\rm V}/g_{\rm A})^2$\\
16 & $H\bm{S}_2$
   & $-\frac{1}{2}[\mu_{\beta}(g_{\rm V}/g_{\rm A}) - 1]$\\
17 & $h\bm{S}_2\bm{y}'_0$ & $-\frac{1}{\sqrt{3}}$\\
18 & $h[\bm{S}_2\otimes\bm{y}'_1]^{(2)}$ & $-\frac{\sqrt{3}}{2}$\\
19 & $h[\bm{S}_2\otimes\bm{y}'_2]^{(2)}$
   & $-\frac{\sqrt{7}}{2\sqrt{3}}$\\
20 & $h[\bm{S}_+\otimes\bm{y}'_1]^{(2)}$
   & $\frac{1}{2\sqrt{2}}(g_{\rm V}/g_{\rm A})$\\
21 & $h[\bm{S}_+\otimes\bm{y}'_2]^{(2)}$
   & $\frac{\sqrt{3}}{2\sqrt{2}}(g_{\rm V}/g_{\rm A})$\\
22 & $h[\bm{S}_1\otimes\bm{Y}'_1]^{(2)}$
   & $-\frac{1}{4}$\\
23 & $h[\bm{S}_1\otimes\bm{Y}'_2]^{(2)}$
   & $-\frac{\sqrt{3}}{4}$\\
24 & $h[\bm{S}_-\otimes\bm{Y}'_1]^{(2)}$
   & $-\frac{1}{4\sqrt{2}}(g_{\rm V}/g_{\rm A})$\\
25 & $h[\bm{S}_-\otimes\bm{Y}'_2]^{(2)}$
   & $-\frac{\sqrt{3}}{4\sqrt{2}}(g_{\rm V}/g_{\rm A})$\\
\hline
\end{tabular}
\end{table}
\begin{table}[h] 
\caption{Calculated matrix elements $M_{\rm m}$ for the $^{76}$Ge and
$^{100}$Mo decays. The entries are the values of the products
$C_{{\rm m}i}M_i$ and their sum $M_{\rm m}$ in units of
$10^{-3}$fm$^{-1}$.}
\medskip
\begin{tabular}{rrrrr}
\hline
$i$ & $^{76}$Ge & $^{100}$Mo\\
\hline
 1 & $-0.0229$ & $-0.0077$\\
 2 &  $0.0013$ &  $0.0003$\\
 3 &  $0.0002$ &  $0.0008$\\
 4 & $-0.0017$ & $-0.0011$\\
 5 & $-0.0003$ &  $0.0007$\\
14 & $-0.0191$ & $-0.0227$\\
15 & $-0.0128$ & $-0.0112$\\
16 & $-0.0033$ & $-0.0006$\\
17 & $-0.0017$ &  $0.0000$\\
18 & $-0.0028$ &  $0.0019$\\
19 & $-0.0005$ &  $0.0001$\\
20 &  $0.0006$ &  $0.0005$\\
21 &  $0.0002$ & $-0.0001$\\
22 &  $0.0020$ & $-0.0000$\\
23 &  $0.0020$ & $-0.0026$\\
24 & $-0.0047$ & $-0.0001$\\
25 &  $0.0012$ & $-0.0010$\\
\hline
sum & $-0.0624$ & $-0.0427$\\
\hline
\end{tabular}
\end{table}
\begin{table}[h] 
\caption{Calculated matrix elements for the $0^+ \to 2^+_1$
$0\nu\beta\beta$ decays of $^{76}$Ge and $^{100}$Mo in units of
$10^{-3}$fm$^{-1}$.}
\medskip
\begin{tabular}{r@{~~~~}l@{~~~~}l@{~~~~}l@{~~~~}l@{}}
\hline
\\[-3ex]
 & ~~~$M_\lambda$
 & ~~~$M_\eta$
 & ~~~$M_\eta'$
 & ~~~$M_{\rm m}$\\
\\[-3ex]
\hline\\[-3ex]
$^{76}$Ge & $\;\;\: 1.81~^{\rm a}$ & ~$13.37~^{\rm a}$
 & ~~$0.18~^{\rm a}$ & $-0.0624$\\
\\[-3ex]
\hline\\[-3ex]
$^{100}$Mo & $-6.33$ & ~$\;\: 3.38$ & ~~5.17 & $-0.0427$\\
\\[-3ex]
\hline
\end{tabular}\par
\medskip
$^{\rm a}$ Ref.\ \cite{TomN}.
\end{table}
\begin{table}[h] 
\caption{Constraints on the right-handed current couplings and the
effective neutrino mass.}
\medskip
\begin{tabular}{cll}
\hline
 & ~~~~~$^{76}$Ge & ~~~~$^{100}$Mo\\
\hline
$|\langle\lambda\rangle|$ & $<8.9\times 10^{-4}$
 & $<3.9\times 10^{-4}$\\
$|\langle\eta\rangle|$ & $<1.2\times 10^{-4}$
 & $<4.3\times 10^{-4}~^{\rm a}$\\
$|\langle m_\nu\rangle|$ [eV] & $<1.0\times 10^{4}$
 & $<2.2\times 10^{4}~^{\rm a}$\\
\hline
\end{tabular}\par
\medskip
$^{\rm a}$ Assuming the same limit on $\tau^{0\nu}_{1/2}$ as the
$\langle\lambda\rangle$ mode.
\end{table}

\begin{thebibliography}{99}
\bibitem[1]{HS} W.C. Haxton and G.J. Stephenson, Jr., Prog. Part. Nucl.
Phys. 12 (1984) 409.
\bibitem[2]{DKT} M. Doi, T. Kotani and E. Takasugi, Prog. Theor. Phys.
Suppl. 83 (1985) 1.
\bibitem[3]{TomR} T. Tomoda, Rep. Prog. Phys. 54 (1991) 53.
\bibitem[4]{Moh} R. Mohapatra, Phys. Rev. D34 (1986) 3457.
\bibitem[5]{Ver} J.D. Vergados, Phys. Lett. B184 (1987) 55.
\bibitem[6]{HKK} M. Hirsch, H.V. Klapdor-Kleingrothaus and
S.G. Kovalenko, Phys. Rev. D53 (1996) 1329.
\bibitem[7]{Mor} A. Morales, Nucl. Phys. B (Proc. Suppl.) 77 (1999) 335.
\bibitem[8]{SV} J. Schechter and J.W.F. Valle, Phys. Rev. D25 (1982)
2951.
\bibitem[9]{Nie} J.F. Nieves, Phys. Lett. 147B (1984) 375.
\bibitem[10]{Tak} E. Takasugi, Phys. Lett. 149B (1984) 372.
\bibitem[11]{TomN} T. Tomoda, Nucl. Phys. A484 (1988) 635.
\bibitem[12]{TFSG} T. Tomoda, A. Faessler, K.W. Schmid and F. Gr\"ummer,
Nucl. Phys. A452 (1986) 591.
\bibitem[13]{SGF} K.W. Schmid, F. Gr\"ummer and A. Faessler, Nucl. Phys.
A431 (1984) 205.
\bibitem[14]{Mai} B. Maier, Nucl. Phys. B (Proc. Suppl.) 35 (1994) 358.
\bibitem[15]{Kud} N. Kudomi {\it et al.}, Nucl. Phys. A629 (1998) 527c.
\bibitem[16]{Bau} L. Baudis {\it et al.}, Phys. Rev. Lett. 83 (1999) 41.
\bibitem[17]{TF} T. Tomoda and A. Faessler, Phys. Lett. B199 (1987) 475.
\end{thebibliography}
\end{document}